\documentclass[aps,prl,twocolumn]{revtex4}
\usepackage{graphicx}
\usepackage{amssymb}
\usepackage[dvips]{color}
\newcommand{\bfk}{{\bf k}}
\newcommand{\ek}{\epsilon_\bfk}
\newcommand{\cks}{c_{\bfk,\sigma}}

\newcommand{\cku}{c_{\bfk,\uparrow}}

\newcommand{\cmkd}{c_{-\bfk,\downarrow}}

\newcommand{\bz}{b_0}

\newcommand{\jm}{{\cal J}_-}
\newcommand{\jp}{{\cal J}_+}
\newcommand{\jx}{{\cal J}_x}
\newcommand{\jy}{{\cal J}_y}
\newcommand{\jz}{{\cal J}_z}

\begin{document}
\title{Nonlinear adiabatic passage from fermion atoms to boson molecules}
\author{E. Pazy$^1$, I Tikhonenkov$^1$, Y. B. Band$^1$, M. Fleischhauer$^2$, 
and A. Vardi$^1$}
\affiliation{$^{1}$Department of Chemistry, Ben-Gurion University of the
Negev, P.O.B. 653, Beer-Sheva 84105, Israel}
\affiliation{$^{2}$Fachbereich Physik, Technische Universit\"at Kaiserslautern, 
D67663, Kaiserslautern, Germany}

\begin{abstract}
We study the dynamics of an adiabatic sweep through a Feshbach resonance
in a quantum gas of fermionic atoms.  Analysis of the dynamical
equations, supported by mean-field and many-body numerical results,
shows that the dependence of the remaining atomic fraction $\Gamma$ on
the sweep rate $\alpha$ varies from exponential Landau-Zener behavior 
for a single pair of particles to a power-law dependence for large 
particle number $N$.  The power-law is linear, $\Gamma \propto \alpha$, 
when the initial molecular fraction is smaller than the $1/N$ quantum 
fluctuations, and $\Gamma \propto \alpha^{1/3}$ when it is larger.  
Experimental data agree better with a linear dependence than with 
an exponential Landau-Zener fit, indicating that many-body effects 
are significant in the atom-molecule conversion process.
\end{abstract}

\maketitle

Adiabatic evolution is an important tool for quantum state
engineering.  The adiabatic theorem ensures that an
initial nondegenerate eigenstate remains an instantaneous eigenstate
when the Hamiltonian changes slowly.  When eigenstates become nearly
degenerate, the Landau-Zener (LZ) model \cite{Landau} is a paradigm for 
explaining how transitions occur.

Adiabatic sweeps across an atom-molecule Feshbach resonance have
recently been used to convert degenerate fermionic atomic gases
containing two different internal spin states to bosonic dimer
molecules \cite{Regal03,Strecker03,Cubizolles03,Hodby}.  Formation of
a molecular condensate has also been observed using both adiabatic
sweeps and three-body recombination processes \cite{condensate}.  In
this Letter we show that for adiabatic Feshbach sweeps that convert
degenerate fermionic atoms to diatomic molecules, the LZ behavior for
a single pair of particles is dramatically changed due to many-body
effects.  The fraction of unconverted atoms is shown to be a power-law
in the sweep rate, rather than exponentially small as predicted by an
essentially single-particle, linear LZ model
\cite{Landau,Mies_Julienne}.  The exact power-law is determined by the
significance of quantum fluctuations.  En route to this result we also
find that, for a ladder of atomic states filled by fermionic atoms,
the atom-molecule sweep efficiency is unaffected by atomic dispersion,
and all fermionic atoms can go over to molecules, in contrast to the
linear LZ model.

We consider the collisionless, interaction representation, single
bosonic mode Hamiltonian
\cite{Javanainen04,Barankov04,Andreev04,Dukelsky04,Tikhonenkov05,%
Miyakawa05,Mackie05}
\begin{eqnarray}
H&=&\sum_{\bfk,\sigma} \ek\cks^\dag\cks+{\cal E}(t)
\bz^\dag\bz\nonumber\\
~&~&+g\left(\sum_{\bf k} \cku\cmkd \bz^\dag+H.c.\right)~,
\label{ham}
\end{eqnarray}
where $\ek=\hbar^2k^2/2m$ is the kinetic energy of an atom with mass
$m$, and $g$ is the atom-molecule coupling strength.  The molecular
energy ${\cal E}(t) = \alpha t$ is linearly swept at a rate $\alpha$
through resonance to induce adiabatic conversion of Fermi atoms to
Bose molecules.  The annihilation operators for the atoms, $\cks$,
obey fermionic anticommutation relations, whereas the molecular
annihilation operator $\bz$ obeys a bosonic commutation relation.

\begin{figure}
\centering
\includegraphics[scale=0.5,angle=0]{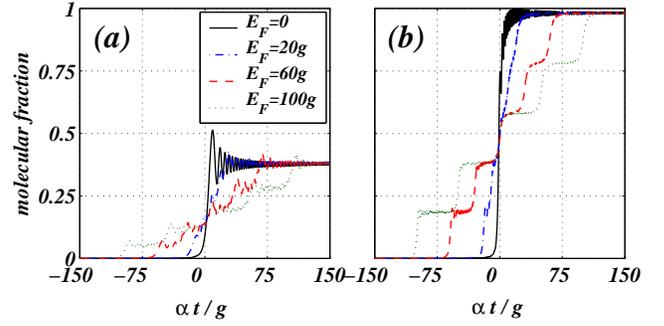}
\caption{Many-body collective dynamics of adiabatic passage from a
fermionic atomic gas into a molecular BEC for five pairs of fermionic
atoms.  (a) Sweep rate $\alpha = 2g^2 N$, (b) Sweep rate $\alpha =
g^2 N/4$.  Overall efficiency is independent of atomic dispersion in
both (a) and (b).}
\label{fig1}
\end{figure}

We find that, provided that all atomic levels are swept through, the
adiabatic conversion efficiency is completely insensitive to the
details of the atomic dispersion.  Fig.~\ref{fig1} shows exact
numerical results for the adiabatic conversion of five atom pairs into
molecules, for different values of the atomic level spacing (and hence
of the Fermi energy $E_F$).  It is evident that, while the exact
dynamics depends on $E_F$, levels are sequentially crossed, leading to
the same final efficiency regardless of the atomic motional
timescale.  In particular, in the limit as $\alpha\rightarrow 0$ it is
possible to convert {\em all} atom pairs into molecules.  This is a
unique feature of the nonlinear parametric coupling between atoms and
molecules, which should be contrasted with a marginal conversion
efficiency expected for linear coupling.  Since the exact energies
$\ek$ do not affect the final fraction of molecules, we use a
degenerate model \cite{Tikhonenkov05,Miyakawa05,Mackie05} with
$\ek=\epsilon$ for all $\bfk$.  In the spirit of
Refs.~\cite{Vardi01,Miyakawa05}, we define the operators:
$$
\jm=\frac{\bz^\dag\sum_{\bfk} \cku\cmkd}{(N/2)^{3/2}} ~,~
\jp=\frac{\sum_{\bfk} \cmkd^\dag\cku^\dag \bz}{(N/2)^{3/2}},
$$
\begin{equation}
\jz=\frac{\sum_{\bfk,\sigma}\cks^\dag\cks-2\bz^\dag\bz}{N},
\label{js}
\end{equation}
where $N=2\bz^\dag\bz+\sum_{\bfk,\sigma}\cks^\dag\cks$ is the
conserved total number of particles. It is important to note that
$\jm,\jp,\jz$ do not span $SU(2)$ as $[\jp,\jm]$ is a quadratic
polynomial in $\jz$.  We also define $\jx=\jp+\jm$ and
$\jy=-i(\jp-\jm)$.  Up to a $c$-number term, Hamiltonian
(\ref{ham}) takes the form
\begin{equation}
H=\frac{N}{2}\left(\Delta(t)\jz+g\sqrt{N\over 2}\jx\right),
\label{hamj}
\end{equation}
where $\Delta(t) = 2\epsilon-{\cal E}(t)$.  Defining a rescaled time
$\tau=\sqrt{N}gt$, we obtain the Heisenberg equations of motion for
the association of a quantum-degenerate gas of fermions,
\begin{eqnarray}
\frac{d}{d\tau}\jx &=& \delta(\tau)\jy \nonumber \\
\frac{d}{d\tau}\jy &=& -\delta(\tau)\jx+\frac{3\sqrt{2}}{4}
\left(\jz-1\right)\left(\jz+\frac{1}{3}\right) \nonumber \\
~&~&-{\sqrt{2}\over{N}}\left(1+\jz\right), \nonumber \\
\label{eom}
\frac{d}{d\tau}\jz&=&\sqrt{2}\jy,
\end{eqnarray}
which depend on the single parameter $\delta(\tau) = \Delta(t)
/\sqrt{N}g = (\alpha/g^2N)\tau$.  We note parenthetically that
precisely the same set of equations, with $\jz\rightarrow -\jz$ and
$g\rightarrow g/2$, is obtained for a two-mode atom-molecule BEC
\cite{Vardi01}, providing another perspective on the recently observed
mapping between the two systems
\cite{Tikhonenkov05,Miyakawa05,Mackie05}.

\begin{figure}
\centering
\includegraphics[scale=0.5,angle=-90]{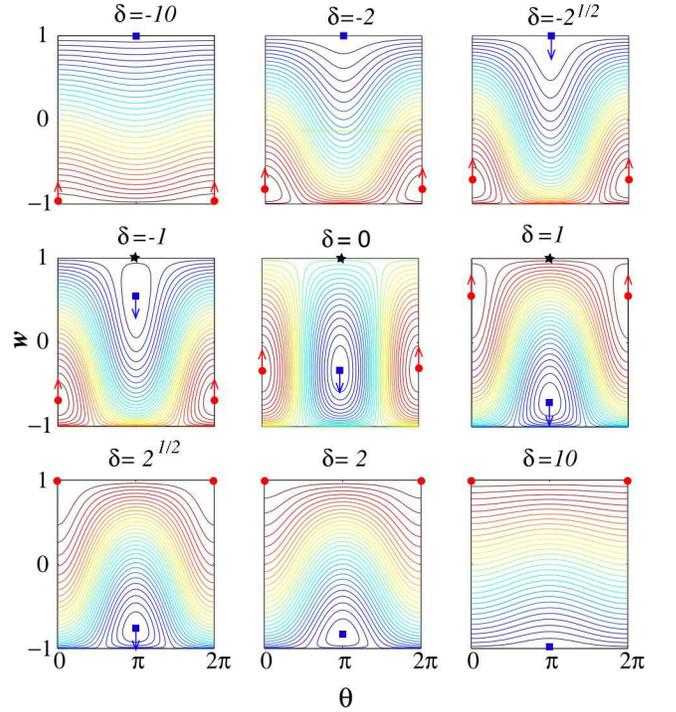}
\caption{Equal-energy contours of Hamiltonian (\ref{hamc}) plotted as
a function of $w$ and $\theta$ for different detunings $\delta$.
$w=1$ is all atoms and $w=-1$ is all molecules.  The various fixed
points corresponding to adiabatic eigenvectors are marked by (blue)
squares, (red) circles and (black) stars.}
\label{fig2}
\end{figure}

We first consider the mean-field limit of Eqs.~(\ref{eom}), replacing
$\jx,\jy$, and $\jz$ by their expectation values $u$, $v$, and $w$
which correspond to the real and imaginary parts of the atom-molecule
coherence and the atom-molecule population imbalance, respectively,
and omitting the quantum noise term $\sqrt{2}(1+\jz)/N$.  In this
limit, the equations depict the motion of a generalized Bloch vector
on a two-dimensional surface, determined by the conservation law,
\begin{equation}
u^2+v^2={1\over 2}(w-1)^2(w+1).
\label{consrv}
\end{equation}
Hamiltonian (\ref{hamj}) is then replaced by the classical form
\begin{equation}
H(w,\theta;\Delta)=\frac{gN^{3/2}}{2}\left(\delta
w+\sqrt{(1+w)(1-w^2)}\cos\theta\right),
\label{hamc}
\end{equation}
with $\theta=\arctan(v/u)$.

To study the atom-molecule adiabatic passage, we closely follow the
method of Ref.~\cite{Liu02}.  The eigenvalues of the atom-molecule
system at any given value of $\delta$ correspond to the fixed points
$(u_0,v_0,w_0)$ of the classical Hamiltonian (\ref{hamc}) or the
mean-field limit of Eqs.~(\ref{eom}):
\begin{equation}
v_0=0~,~\frac{\sqrt{2}}{4}\left(w_0-1\right)\left(3w_0+1\right)=\delta
u_0.
\label{fxp}
\end{equation}
The number of fixed points depends on the parameter $\delta$.  The
point $u_0=v_0=0,w_0=1$ is stationary for any value of $\delta$.
Using Eqs.~(\ref{consrv}) and (\ref{fxp}), other fixed points satisfy
\begin{equation}
\frac{(3w_0+1)^2}{4(w_0+1)}=\delta^2.
\label{roots}
\end{equation}
In Fig.~\ref{fig2} we plot phase-space trajectories, corresponding to
equal-energy contours of Hamiltonian (\ref{hamc}), for different
values of $\delta$.  As expected from (\ref{hamc}), the plots have the
symmetry $(w,\theta;\delta) \leftrightarrow (w,\theta+\pi;-\delta)$.
For sufficiently large detuning, $|\delta|> \sqrt{2}$,
Eq.~(\ref{roots}) has only one solution in the range $-1\leq w_0\leq
1$.  Therefore, there are only two (elliptic) fixed points, denoted by
a red circle corresponding to the solution of Eq.~(\ref{roots}), and a
blue square at (0,0,1).  As the detuning is changed, one of these
fixed points (red circle) smoothly moves from all-molecules towards
the atomic mode.  At detuning $\delta=-\sqrt{2}$ a homoclinic orbit
appears through the point $(0,0,1)$ which bifurcates into an unstable
(hyperbolic) fixed point (black star) remaining on the atomic mode,
and an elliptic fixed point (blue square) which starts moving towards
the molecular mode.  Consequently, in the regime $|\delta|<\sqrt{2}$
there are two elliptic fixed points and one hyperbolic fixed point,
corresponding to the unstable all-atoms mode.  Another crossing occurs
at $\delta=\sqrt{2}$ when the fixed point which started near the
molecular mode (red circle) coalesces with the all-atoms mode (black
star).

The frequency of small periodic orbits around the fixed points,
$\Omega_0$, is found by linearization of the dynamical equations
(\ref{eom}) about $(u_0,v_0,w_0)$ and using (\ref{roots}) to obtain
\begin{equation}
\frac{\Omega_0}{g\sqrt{N}}=\sqrt{\delta^2+(1-3w_0)} =
\sqrt{\frac{(1-w_0)(3w_0+5)}{4(w_0+1)}}~.
\label{omegz}
\end{equation}
Hence, for $|\delta|<\sqrt{2}$ the period of the homoclinic trajectory
beginning at $(0,0,1)$ diverges.

Transforming $w,\theta$ into action-angle variables $I,\phi$, the
non-adiabatic tunneling probability $\Gamma$ at any finite sweep rate
$\alpha$ is related to the action $I$ accumulated during the sweep
\cite{Landau,Liu02,Garraway00},
\begin{equation}
\Gamma^2=\frac{\Delta I}{2}={1\over 2}\int_{-\infty}^{\infty}
R(I,\phi) \, {\dot\Delta} \, \frac{d\phi}{{\dot\phi}}~,
\label{nonadiab}
\end{equation}
where $R(I,\phi)$ is related to the generating function of the
canonical transformation $w,\theta\rightarrow I,\phi$.  We note that,
unlike the linear \cite{Landau} or Josephson \cite{Liu02,Garraway00}
case, where the tunneling probability is linearly proportional to the
action increment $\Delta I$, our choice of variables (\ref{js}) causes
the atomic population at the end of the sweep (and hence, $\Gamma$) to
be proportional to the {\it square root} of $\Delta I$ (since
$u^2(t_f)+v^2(t_f)\propto\left|\sum_{\bfk,\sigma}
n_{\bfk,\sigma}(t_f)\right|^2$, where $n_{\bfk,\sigma}(t_f)$ is the
population in state $|\bfk,\sigma\rangle$ at the final time $t_f$).
Equation (\ref{nonadiab}) depicts the familiar rule that in order to
attain adiabaticity, the rate of change of the adiabatic fixed points
through the variation of the adiabatic parameter $\Delta$,
$R(I,\phi)\, {\dot\Delta}$, should be slow with respect to the
characteristic precession frequency ${\dot\phi}=\Omega_0$ about these
stationary vectors.  For an adiabatic process where
${\dot\Delta}/{\dot\phi}\rightarrow 0$, the action (which is
proportional to the surface-area enclosed within the periodic orbit)
is an adiabatic invariant, so a zero-action elliptic fixed point
evolves into a similar point trajectory.  Action is accumulated mainly
in the vicinity of singularities where ${\dot\phi} =
\Omega_0\rightarrow 0$.  For linear adiabatic passage \cite{Landau},
such singular points lie exclusively off the real axis, leading to
exponential LZ transition probabilities.  However, when nonlinearities
are dominant, as in the Mott-insulating Josephson case
\cite{Liu02,Garraway00} and our case, there are real singularities,
leading to power-law dependence of the transfer efficiency on the
sweep rate.

It is clear from Eq.~(\ref{omegz}) that, for the atom-molecule
conversion problem, a real singularity of the integrand in
Eq.~(\ref{nonadiab}) exists at $w_0=1$, where the frequency vanishes
as $\Omega_0\approx g\sqrt{N(1-w_0)}$.  Thus, most the the
nonadiabatic correction is accumulated in the vicinity of this point
(all-atoms for fermions and all-molecules for bosons).  Taking the
derivative of Eq.~(\ref{roots}) with respect to time, we find that the
response of the fixed-point velocity to a linear sweep rate $\alpha$
is,
\begin{equation}
{{\dot w}_0}=\frac{4\alpha}{g\sqrt{N}}\frac{(w_0+1)^{3/2}}{3w_0+5}.
\label{wdot}
\end{equation}
Having found ${{\dot w}_0}$, we can now find the action-angle variable
$\phi$ in terms of $w_0$: $\phi=\int {\dot\phi} dt =\int \Omega_0
\frac{dw_0}{{\dot w}_0}$.  In the vicinity of the singularity we have
$\Omega_0 \approx g\sqrt{N(1-w_0)}$ and ${\dot w}_0\approx
{\sqrt{2}\alpha/g\sqrt{N}}$, resulting in
\begin{equation}
\phi=\frac{g^2 N}{\alpha}\frac{\sqrt{2}}{3}(1-w_0)^{3/2}.
\label{phiofw}
\end{equation}
Using Eq.~(\ref{phiofw}), we finally find that near the singularity,
$\dot\phi=\Omega_0 \approx g\sqrt{N(1-w_0)}$ is given in terms of
$\phi$ as
\begin{equation}
{\dot\phi}=\left(3\sqrt{N\over 2} g\alpha\right)^{1/3}\phi^{1/3}.
\label{dotphiofphi}
\end{equation}
Substituting (\ref{dotphiofphi}) and $\dot\Delta=\alpha$ into Eq.
(\ref{nonadiab}) we find that the nonadiabatic correction  depends on
$\alpha$ as
\begin{equation}
\Gamma \propto \alpha^{1/3}~.
\label{cubicroot}
\end{equation}

\begin{figure}
\centering
\includegraphics[scale=0.5,angle=0]{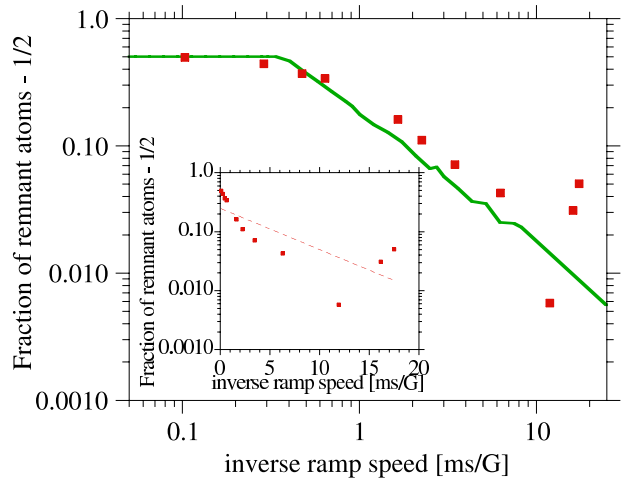}
\caption{Fraction of remnant atoms versus inverse ramp speed
$1/\dot{B}$ across the 543 G resonance in a two-component Fermi gas of
$^6$Li.  The experimental data (red squares) of
Ref.~\cite{Strecker03}, which saturates at a remnant of $1/2$
\cite{Pazy04}, and the mean-field calculations obey a linear
dependence on sweep rate beyond 0.4 ms/G. ${g^2\over\alpha N}$ is
multiplied by 0.4 ms/G to scale the abscissa for the calculated
results.  The insert shows the best exponential fit of the data as a
dashed line.}
\label{fig3}
\end{figure}

So far, we have neglected the effect of quantum fluctuations, which
are partially accounted for by the source term $(\sqrt{2}/N)(1+\jz)$
in Eqs.~(\ref{eom}).  As a result, we found that $\dot w_0$ does not
vanish as $w_0$ approaches 1.  Consequently, the remaining atomic
population is expected to scale as the cubic root of the sweep rate if
the initial average molecular fraction is larger than the quantum
noise.  However, starting purely with fermion atoms (or with molecules
made of bosonic atoms), fluctuations will initially dominate the
conversion process.  Equation (\ref{roots}) should then be replaced by
\begin{equation}
\delta=\frac{2}{\sqrt{w_0+1}}\left(\frac{3w_0+1}{4}-\frac{w_0+1}{N(w_0-1)} 
\right)~,
\end{equation}
demonstrating that our previous treatment around $w_0=1$ is only
valid provided that $|w_0(t_i)-1|\gg 1/N$. For smaller initial
molecular population, Eq.~(\ref{wdot}) should be replaced by
\begin{equation}
{{\dot w}_0}=\frac{\alpha}{g\sqrt{N}}\left/\left[
\frac{3w_0+5}{4(w_0+1)^{3/2}}+\frac{w_0+3}{N(w_0+1)^{1/2}(w_0-1)^2}
\right]\right. .
\end{equation}
Hence, in the vicinity of $w_0=1$ the eigenvector velocity in the $w$
direction vanishes as ${{\dot w}_0} = (\sqrt{N}\alpha/g\sqrt{8})
\left(w_0-1\right)^2$.  The characteristic frequency $\dot\phi$ is now
proportional to $(\alpha\phi)^{-1}$ instead of Eq.~(\ref{dotphiofphi})
so that $\Delta I\propto\alpha^2$, and \cite{Ishkhanyan04,Altman05}
\begin{equation}
\Gamma \propto \alpha~.
\label{linear}
\end{equation}

Equations (\ref{linear}) and (\ref{cubicroot}) constitute the main
results of this work.  We predict that the remnant atomic fraction in
adiabatic Feshbach sweep experiments will scale as a power-law with
sweep rate due to the curve crossing in the nonlinear case.  The
dependence is expected to be linear if the initial molecular
population is below the quantum-noise level (i.e., when $1-w_0(t_i)\ll
1/N$), and cubic-root when fluctuations can be neglected (i.e. for
$1-w_0(t_i) \gg 1/N$).  We note that a similar linear dependence was
predicted for adiabatic passage from bosonic atoms into a molecular
BEC \cite{Ishkhanyan04}.

\begin{figure}
\centering
\includegraphics[scale=0.5,angle=-90]{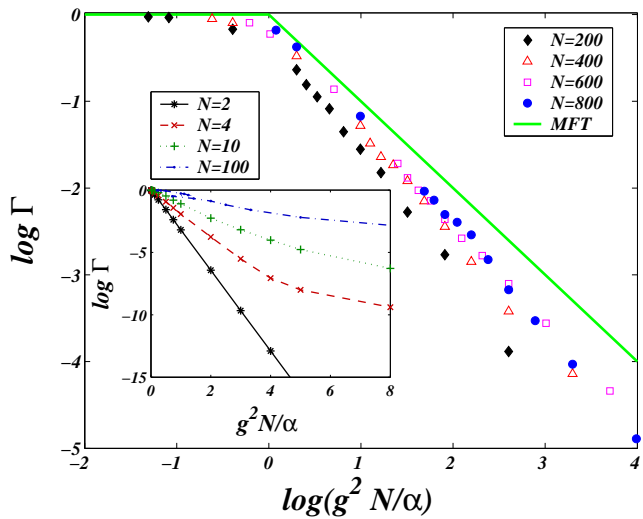}
\caption{Many-body calculations for the fraction of remnant atoms
versus dimensionless inverse sweep rate for various particle numbers
in the range $N = 2$ to 800.  The many body results for large number 
of particles converge to the mean-field results (solid green line) 
of Fig.~\ref{fig3}.}
\label{fig4}
\end{figure}

The analytical predictions illustrated above are confirmed by
numerical simulations and existing experimental data.  Fig.~\ref{fig3}
shows the fraction of remnant atoms $\Gamma$ versus inverse sweep rate
${g^2\over\alpha N}$, computed numerically from the mean-field limit
of Eqs.~(\ref{eom}).  The log-log plot highlights the power law
dependence obtained beyond 0.4 ms/G (i.e., for ${g^2\over\alpha N} >
1$).  The numerical results compare quite well with the experimental
data of Ref.~\cite{Strecker03} (red squares in Fig.~\ref{fig3}),
indicating that nonlinear effects do indeed dominate the adiabatic
passage dynamics in these experiments.  The power-law fit of the
experimental results is contrasted with an exponential LZ fit (insert
of Fig.~\ref{fig3}) which fails to provide an accurate description of
the observed dependence of efficiency on sweep rate.

To go beyond the mean-field analysis we carry out exact many-body
calculations of efficiency vs.~dimensionless inverse sweep-rate,
${g^2\over\alpha N}$, using the methodology of \cite{Tikhonenkov05}.
The results are shown in Fig.~\ref{fig4}.  For a single pair of
particles, $N=2$, the quantum association problem is formally
identical to the linear LZ paradigm, leading to an exponential
dependence of the remnant atomic fraction on sweep rate (see insert of
Fig.~\ref{fig4}).  However, as the number of particles increases,
many-body effects come into play, and there is a smooth transition to
a power-law behavior in the regime $\alpha<g^2 N$.  We note that this
is precisely the regime where Eq.~(\ref{nonadiab}) can be used to
estimate $\Delta I$ and $\Gamma$ \cite{Landau}.

In summary, we have shown that nonlinear many-body effects play a
significant role in the atom-molecule conversion process for
degenerate fermionic atomic gases, modifying the LZ exponential
dependence on sweep rate.  Experimental data seem to back up this
conclusion.

\begin{acknowledgments}
This work was supported in part by grants from the U.S.-Israel
Binational Science Foundation (grant Nos.~2002214, 2002147), the
Minerva Foundation through a grant for a Minerva Junior Research
Group, the Israel Science Foundation for a Center of Excellence (grant
No.~8006/03), and the German Federal Ministry of Education and
Research (BMBF) through the DIP project.
\end{acknowledgments}

\end{document}